\documentclass[prd,
nofootinbib, showpacs, twocolumn,
]
{revtex4}
\usepackage{graphicx}

\def\NGB{Nambu-Goldstone boson }
\def\NG{Nambu-Goldstone }

\def\hbar{\hspace{0pt}\raisebox{1pt}{$-$} \hspace{-7pt} h}

\def\5{\overline 5}

\newcommand{\be}{\begin{equation}}
\newcommand{\ee}{\end{equation}}
\newcommand{\ba}{\begin{eqnarray}}
\newcommand{\ea}{\end{eqnarray}}
\newcommand{\no}{\nonumber}

\begin{document}
\title[Massive resonant modes as the relevant excitations
of gravitating branes] {Massive states as the relevant deformations
of gravitating branes}
\author{Luca~Vecchi}

\affiliation{INFN, Sezione di Trieste and\\
Scuola Internazionale Superiore di Studi Avanzati (SISSA/ISAS)}
\begin{abstract}
Five-dimensional theories manifesting spontaneous
brane generation are discussed
in a gravitational context. Without gravity, the IR dynamics of the brane fluctuation below
the brane tension scale is described by
an effective theory for the \NG bosons. When gravity is properly taken into account the long distance dynamics changes. The spontaneous breaking of local translational invariance triggers the formation of massive representations via the Higgs mechanism and induces the appearance of new mass scales in the IR. These scales can in principle depend on other fundamental parameters besides the brane tension and the Planck scale. In noncompact extra dimensions the massive states are found to be scalar bound states. We obtain explicit expressions for their propagator and show that their masses depend on the brane width and are thus much heavier than expected. We present an exactly solvable model which captures the main features of the gravitational system.
 \noindent
\end{abstract}
\pacs{04.50.+h}
\maketitle
%

\section{Introduction and results}

Spontaneous breaking of continuous symmetries predicts the
existence of \NG modes as relevant field coordinates for the low
energy dynamics. If the symmetry is gauged a massive
representation emerges in place of them due to the Higgs
phenomenon. The mass scale generated by this mechanism may
represent an important threshold for the IR theory, as the Fermi
coupling in the weak interactions.

An analogous consideration holds for the case of spontaneous breaking
of translational invariance. If the spontaneous breaking refers to
a compact coordinate a Higgs phenomenon is expected to involve the
graviphoton~\cite{ADD}. A rigorous analysis of the nonlinear
realization of the local space-time symmetries shows that the \NG boson
kinetic term (the Nambu-Goto action) does not provide any mass
term for the graviphoton~\cite{Love}, as naively expected by
analogy with internal gauge theories. The mass term arises from
additional operators that one can build out of the relevant IR
variables. These operators are essential for the self-consistency of the
description since they encode the presence of the symmetry
breaking, and thus ensure continuity of the observable as gravity is
switched off. In this phenomenological approach the mass of the
graviphotons cannot be rigorously connected to other fundamental
scales of the theory (such as the brane tension and the 4d Planck
mass) and remains a free parameter. Its determination is crucial for a realistic study of the phenomenology of these models.

The main purpose of the paper is to analyze the emerging of this
gravity-induced scale in an exactly tractable framework. We will
study a class of gravitational models in which a scalar field
develops a nontrivial background along a single space-like
direction. Without loss of generality we consider the 5d
lagrangian \ba\label{act} {\cal L} = \sqrt{-g}\left(
-M^{3}R+g^{AB}\partial_A\Phi\partial_B\Phi-V(\Phi)+\dots\right)
\label{lag} \ea where the dots refer to some unspecified theory
coupled to the scalar $\Phi$ (and gravity). The $5$-dimensional
coordinates are denoted as $x^\mu,y$, with $A,B=0,1,2,3,5$, Greek
indeces referring to the unbroken coordinates, while $y$ denotes
the broken spacial direction. The metric signature is "mostly
minus".

The $y$-dependent vacuum expectation value
$\langle\Phi\rangle\equiv\Phi_0(y)$ is assumed to trigger the
localization of light modes of the above unspecified theory around
some point of the $y$-direction, which we conventionally choose as
the origin. This effectively realizes a brane world with a
nontrivial dynamics trapped on it. In order to keep our
discussion as general as possible we will not write down any
explicit function $\Phi_0$, nor any action for the would-be
localized fields. We simply stress that natural candidates for
$\Phi_0$ are the kink shape proposed in~\cite{RubShap} (in this
case the dots in~(\ref{act}) would stand for fermionic fields), or
one of the scalars described in~\cite{DS} (in this case the dots
in~(\ref{act}) would stand for gauge fields). For simplicity the background geometry preserves $4d$-Poincare
invariance \ba\label{sol} ds^2=a^2\eta_{\mu\nu}dx^\mu dx^\nu-dy^2,
\ea where $a=e^{A}$ and $A=A(y)$. 

The equations of motion derived
from~(\ref{act}) reduce to the following independent conditions
\ba\label{EOM} \dot\Phi_0^2&=&-3M^3\ddot A
\\\no V(\Phi_0)&=&-3M^3(4\dot A^2+\ddot A),
\ea where $\dot f\equiv df/dy$. We see that the first
equation~(\ref{EOM}) requires $\ddot A\leq0$, this being a general
consequence of the energy conditions~\cite{WEC}. Such a
constraint implies that no regular solution of the equations of
motion is admitted on a circle (we are restricting our analysis to
two-derivatives theories). Being interested in the study of
translationally invariant theories, we are forced to consider a
noncompact extra dimension $y$. This, by itself, may not represent a serious problem for our study
because the authors of~\cite{RS} succeeded in making sense of
these theories, at least for what concerns the spin-2
sector\footnote{These theories are expected to provide a sensible
effective description of the brane physics even though an apparent
inconsistency arises in the evaluation of the
self-couplings~\cite{Porrati,RS}.}. Similarly, we will be able to
obtain an effective description of the brane fluctuations. Another
feature of noncompact extra dimensions is more subtle: the
graviphotons are not dynamical fields. This raises up another
question: what about the gravity-induced scale if the Higgs
mechanism is not canonically realized? An answer to this question
is provided by the analysis of~\cite{STZ}. These authors studied
in detail the scalar kink background implementing the light-cone gauge and showed that continuity of long range potentials between $\Phi$ sources is ensured by the presence of a massive
resonance in the scalar sector.

We will identify gauge-independent properties of this resonance and show 
that the main features of the coupled scalar-gravity system can be
captured by an exactly solvable model. Under some simplifying assumptions (in this set up
the requirement $k<w$, where $k$ is the curvature scale and $w$ is
the inverse brane width, is not an option), we will be able to
extract an explicit expression for the resonant propagator. The resonance
couplings to brane localized currents are those of the zero mode
of the global theory, except for corrections of order O$(k/w)$. In
the realistic limit $k\ll w$, and at energies below the scale $w$,
the resonance represents the only relevant deformation of the
brane, very much like the zero mode of the gravity-free model.

In this class of models the resonance mass $m_R$ plays an analogous
role as the graviphoton mass in compact extra dimensions: for
energies much larger than $m_R$ one recovers the \NGB dynamics via
a generalization of the equivalence theorem while, at scales much smaller, the information
about the spontaneous brane generation is encoded in
nonrenormalizable operators suppressed by the scale $m_R$.

In the following we will analyze in detail the physics of the
brane fluctuations (the $\Phi$ excitations) and their relevance on
the brane-localized dynamics, with and without gravity. Even
though the class of backgrounds considered here are to some extent
special, as we have seen, we believe it can shed light on the
effect of the gauging of space-time symmetries in more general
scenarios.

\section{The linearized theory without gravity}

In order to render the paper self-consistent we review some of the
basic properties of the nongravitational theory. Without loss of
generality we consider the potential $V=(\delta_\Phi W)^2$, with
$W$ an arbitrary function of $\Phi$ and a $\delta_\Phi$ indicating
derivative with respect to the field. The system admits degenerate
constant vacua which, by convention, have zero energy. These
solutions have $\delta_\Phi W=0$, and consequently
$\delta_\Phi^2V=2(\delta_\Phi^2W)^2\geq0$, while nontrivial
solutions satisfy $\dot\Phi_0=\delta_\Phi W$.

We will quantize our theory on the background $\Phi_0(y)$ which,
as anticipated in the introduction, is supposed to trigger the
formation of a brane at $y=0$. A prototypical example which will
be used as a reference is the kink background \ba\label{kink}
\Phi_0=v\tanh(wy) \ea which follows from \ba
W=wv\Phi\left(1-\frac{\Phi^2}{3v^2}\right). \ea

The spectrum of the field $\Phi$ is obtained by studying the
linearized equation for the fluctuation $\phi\equiv\Phi-\Phi_0$,
which is easily found to be \ba\label{phi}
\left(\partial_\mu\partial^\mu-\partial_y^2+\frac{\dddot\Phi_0}{\dot\Phi_0}\right)\phi=0,\ea
where, because of the $y$ dependence of the background, the relation
$\delta_\Phi^2V(\Phi_0)=2\dddot\Phi_0/\dot\Phi_0$ holds. The
eigenmodes $\phi_m(y)$ are solutions of eq.~(\ref{phi}) provided
we replace $\partial_\mu\partial^\mu\rightarrow -m^2$. They form a
complete set in the space of square integrable functions and can
be used to expand in Kaluza-Klein modes the 5d field
$\phi(x,y)=\sum_m \phi_m(y)Q_m(x)$ and write an effective 4d
lagrangian: \ba\label{eff} {\cal L}_{4d}=\sum_m N_m (\partial_\mu
Q_m\partial^\mu Q_m-m^2Q_m^2), \ea where $N_m=\int dy\,\phi_m^2$.

The spectrum of the brane fluctuation generally contains a zero
mode $Q_0$ with wavefunction $\phi_0\propto\dot\Phi_0$ (which we
take to be normalizable), possible discrete eigenvalues, and a
continuum starting at a threshold defined by the potential
$\delta_\Phi^2V$. The interacting terms are obtained as usual from
the convolutions of the 5d profiles and, after the fields $Q_m$
have been properly normalized, they turn out to depend on inverse
powers of the normalizations $N_m$.

An important comment is in order. In the absence of gravity the
system is truly translational invariant. A global shift
$y\rightarrow y+\xi$ changes the nontrivial background
configuration $\Phi_0(y)$ into a new vacuum solution. Promoting
the parameter to a 4d field $\xi(x)$ we can identify it with
the \NG mode. Although the Lorentz rotations orthogonal to $y$ are
spontaneously broken, as well, they do not act independently on
the vacuum~\cite{Low} so that a single massless mode is predicted.
The \NGB cannot be found as a dynamical
mode in the linearized approach. Indeed, even though at infinitesimal level in the
symmetry transformation it coincides with the zero
mode $Q_0(x)$, ensuring its mass is exactly zero, at a
nonlinear level all of the $Q_m$ have a
nontrivial overlap with it. Hence the fields $Q_m$ are seen to
acquire a potential, while the \NGB interacts only
via derivative couplings. The strength of these couplings
follows from the \NGB normalization, which also coincides with
$N_0$, \ba \int dy\, \dot\Phi_0^2=\int dy\,
\partial_yW=\Delta W, \ea where we defined the topological charge
$\Delta W=W(+\infty)-W(-\infty)$. This quantity measures the
energy density of the dimensionally reduced system: \ba
\rho[\Phi_0]=\int dy \left({\dot\Phi_0}^2+V\right)=2\Delta W, \ea
i.e. the brane tension.

A description of the dimensionally reduced theory can be obtained
using an effective approach that makes no reference to the physics
responsible for the generation of the defect~\cite{Sundrum}. This
is a very powerful approach if we ignore the dynamics responsible
for the brane generation, but it cannot tell much about the
relevance of the excitations ignored in the description. Since we
have at our disposal an explicit model, an effective description
in terms of a \NGB is not convenient and we will adopt the
language of the linearized theory of the fluctuations. Making
contact between these two approaches is not immediate because the
\NGB is a composite state of the fields $Q_m$.

We will focus on the brane dynamics at energies much smaller than
the characteristic mass of the modes $Q_m$. This scale also
defines the inverse brane width $w$, so that an observer at
momentum $p^2\ll w^2$ would see the brane as an infinitely thin
defect. Consequently we call this regime the thin brane limit. In
this regime the mode $Q_0$ is the only accessible dynamical
excitation of the brane, the massive states being integrated
out\footnote{Because of the composite nature of the \NGB the
states $Q_m$, for $m\neq0$, cannot decouple in the limit
$w\rightarrow\infty$. Their integration leads to corrections to
the $Q_0$ potential which are of the same order as the bare
couplings.}. The low energy dynamics on the brane is thus
described by an effective theory of the $Q_0$ mode coupled
(generically via nonderivative operators) to some
(unspecified) physics localized at $y=0$.

Before embarking on the study of the implications of the gauging
of the translational symmetry on this effective theory, it is
worth to anticipate some tools that will be used in the following
sections.

When gravity is switched on, eq.~(\ref{phi}) gets extremely
involved and it will not be easy to extract useful information
from it. It is therefore convenient to develop a systematic method
to approximate the eigenvalue problem. Let us illustrate how this
works in flat space-time. First of all we introduce the following
parametrization $\phi_m=\dot\Phi_0 f_m$ so that the eigenvalue
equation reads \ba -\ddot f_m-2\frac{\ddot\Phi_0}{\dot\Phi_0}\dot
f_m=m^2f_m.\ea The convenience of this parametrization is that we
extracted the zero mode profile and we can now simplify the
eigenvalue problem without loosing the main dynamical ingredient.
In order to achieve this it is necessary to make some assumptions
on the background configuration. We will assume that the defect is
exponentially localized as $\dot\Phi_0\sim e^{-2w|y|}$, i.e. the
zero mode is normalizable, and that $\ddot\Phi_0(0)$ is odd, i.e.
the zero mode has a sharp peak on the brane. Under our assumptions
we can write
 \ba -\ddot
f_m+4w\,sign(y)\dot f_m=m^2f_m\ea from which it follows, except
for a normalization, the general solution \ba
f_m=e^{2w|y|}\left(\cos(\mu |y|)+\beta_m\sin(\mu |y|)\right)\ea
with $\mu^2=m^2-4w^2$. The boundary condition on the brane $\dot
f(0)=0$ and normalizability fully determine the spectrum. The
simplified problem clearly predicts a zero mode $f_0=const$, and a
continuum $m>2w$ of delta function normalizable
modes\footnote{Under our assumptions, a possible next to higher
localized mode would have odd parity and would not be crucial for
the effective brane dynamics.}.

For later convenience we also define the brane to brane propagator
as the $\phi$ two-point function $G(p^2,y,y')$ in 4d momentum
space and for $y=y'=0$. From its formal definition it follows
$G(p^2,y,0)\propto\phi_p(y)$ for any $y\neq0$. Since $\phi_m$ is
smooth everywhere and satisfies trivial boundary conditions at the
origin, the normalization reads $\dot G(p^2,0,0)=1$. This
specifies the solution \ba
G(p^2,y,0)=\frac{\phi_p(y)}{\dot\phi_p(0)}\ea up to boundary
conditions in the asymptotic region $|y|\rightarrow\infty$. We can
isolate the poles of the discrete spectrum by requiring an
asymptotic exponential behavior for the Green's function. This is
done by imposing $\beta_p=i$ in the approximate expressions
obtained above. Using our parity assumption ($\ddot\Phi_0(0)=0$) and expanding in $p/w$ keeping the dominant contribution only, we see that the brane to brane correlator acquires
a pole at $p^2=0$: \ba\label{propNGB}
G(p^2,0,0)\sim\frac{4w}{p^2}.\ea The pole appears in the real axis
and thus corresponds to a physical massless particle. The residue
$4w\sim\dot\Phi_0^2(0)/\int dy\, \dot\Phi_0^2$ sets the strength
of a typical interaction between the zero mode and arbitrary
brane-localized currents.

\section{The linearized scalar sector with gravity}

Smooth scalar backgrounds triggering brane generation can be found
in a gravitational context in terms of a function $W$~\cite{K}.
The solutions for~(\ref{EOM}) read $\dot\Phi_0=\delta_\Phi W$ and
$-3M^3\dot A=W $ provided \ba V&=&\left(\frac{\delta
W}{\delta\Phi}\right)^2-\frac{4}{3}\frac{W^2}{M^3}.\ea For
example, the kink background is defined by the same $W$ introduced
in the previous section. Its backreaction gives rise to a warp
factor (assuming $A(y)$ even and choosing the normalization
$A(0)=0$) \ba A=-\frac{2}{9}\frac{v^2}{M^3}
\left(\log\cosh(wy)+\frac{1}{4}\tanh(wy)^2\right). \ea We can
interpret this solution as a smooth realization of the
Randall-Sundrum geometry. If we explore energies of order $p^2\ll
w^2$, the brane looks like an infinitely thin defect and the warp
factor can be approximated as $A=-k|y|$, with \ba\label{k}
k=\frac{2}{9}\frac{v^2}{M^3}w=\frac{\Delta W}{M^3}.\ea The second
equality holds up to numerical factors and it is completely
general. The relation $k\ll w$ is thus forced by the requirement
of a sensible semiclassical approach to gravity.

When translational symmetry is made local a shift of the vacuum
along the broken direction no longer identifies a dynamical
perturbation. As anticipated in the introduction, no Higgs
mechanism is expected since the graviphotons are found to be
unphysical perturbations in these backgrounds. The spontaneous
breaking will modify the pure scalar sector of the gravitational
theory. The analysis of the latter has been carried out in detail
in~\cite{Gio,peloso} (for a generalization to nonminimal
couplings see~\cite{AV}). Let us review the main conclusions and,
for convenience, express the results in the conformal coordinate
$z$, $dy=adz$.

The nontrivial background induces a mixing between the field
excitation $\phi=\Phi-\Phi_0$ and the scalar components of the
metric. In the longitudinal gauge and at linearized level the
perturbed line element reads \ba\label{metric}
ds^2=a^2[(1+F)\eta_{\mu\nu}dx^\mu dx^\nu-(1-S)dz^2]. \ea The
additional $\delta g_{AB}$ components include tensor and vector
fields which play no role in the diagonalization of the scalar
sector. They give rise to a continuous spectrum of massive spin-2
fields, and a normalizable zero mode identified with the graviton.
We refer the reader to the literature for details on the spin-2
dynamics, in the following we will discuss the pure spin-0 sector.

The scalar sector is subject to two constraints. In the
longitudinal gauge the first requires $S=2F$, while the second is
the formal statement that the fields $F$ and $\phi$ are not
independent fluctuations. For arbitrary gauge choices this reads:
\ba\label{const} a^2(F'+A'
S)=-\frac{2}{3}\frac{1}{M^3}a^2\Phi_0'\phi, \ea where a prime
indicates derivative with respect to $z$. We conclude that the
system admits a single independent 5d scalar fluctuation.

The fields $F$ and $\phi$ satisfy two dynamical equations in
addition to the constraint~(\ref{const}). These are more elegantly
expressed in terms of gauge invariant variables~\cite{Gio} as:
\ba\label{dyn} ({\cal A}^+{\cal
A}^-+\partial_\mu\partial^\mu){\cal G}&=&0\\\no ({\cal A}^-{\cal
A}^++\partial_\mu\partial^\mu){\cal U}&=&0\ea where \ba {\cal
A}^{\pm}=\pm\partial_z+\frac{{\cal Z}'}{{\cal Z}},\quad\quad {\cal
Z}=a^{3/2}\frac{\Phi_0'}{A'},\ea and the diffeomorphism invariant
variables are defined in the longitudinal gauge as \ba\label{gi} {\cal
G}&=&a^{3/2}(\phi-\frac{\Phi_0'}{2A'}F)\\\no {\cal U}&=&
a^{3/2}\frac{F}{\Phi_0'}.\ea

In order to analyze the spectrum we decompose the 5d fields
$F,\phi$ in Kaluza-Klein modes $F(x,z)=\sum_m F_m(z)Q_m(x)$ and
$\phi(x,z)=\sum_m\phi_m(z)Q_m(x)$, where the eigenfunctions
$F_m(z),\phi_m(z)$ satisfy eqs.~(\ref{const}) and~(\ref{dyn}) with
$\partial_\mu\partial^\mu\rightarrow -m^2$. Because
of~(\ref{const}) the equations~(\ref{dyn}) are equivalent for $m\neq0$. More precisely, for any eigenvalue $m$
(including $m=0$) the eigenfunctions are related by \ba\label{0}
{\cal A}^-{\cal G}_m=-\frac{3M^3}{2}m^2{\cal U}_m.\ea

In order to complete the eigenvalue problem for the modes
$F_m,\phi_m$ and determine the spectrum we need to specify the
normalization condition. The 4d kinetic terms for the scalars
$F$,$\phi$, and $S$ in the longitudinal gauge
is~\cite{Gio,peloso,AV} \ba
a^3\left(\partial_\mu\phi\partial^\mu\phi-\frac{3}{2}M^3\partial_\mu
F(\partial^\mu F-\partial^\mu S)\right). \ea Substituting the
condition $S=2F$ we can immediately derive the normalization of
the dynamical fields $Q_m$~\cite{peloso}: \ba\label{N} N_m&=&\int
dz\,a^3\left[\phi_m^2+\frac{3}{2}M^3F_m^2\right]\\\no
&=&\frac{3M^3}{2}\int
\frac{dy}{a^2}\left[\frac{3}{2}\frac{M^3}{\dot\Phi_0^2}\dot
Y_m^2+Y_m^2\right], \ea where in the second equality we changed
back to the coordinate $y$. In the last expression the definition
$Y_m\equiv a^2F_m$ and the constraint~(\ref{const}), i.e. $\dot
Y_m=-2a^2\dot\Phi_0\phi_m/3$, have been used.

The authors~\cite{peloso} have proven the hermiticity of the
quadratic action in the fluctuations under very general boundary
conditions. This ensures that the Kaluza-Klein expansion is
meaningful. The 4d dependent coefficients $Q_m(x)$ play the role
of the physical 4d fields and satisfy the dispersion relation
$(\partial_\mu\partial^\mu+m^2)Q_m=0$. One can thus write a free
lagrangian formally equivalent to~(\ref{eff}). The crucial
difference between these two theories is the spectrum of physical
states.

The physical spectrum of the perturbations, including $m=0$, is
specified by the second of the equations~(\ref{dyn}) together with
the requirement $N_m<\infty$. Since the normalization condition
can be conveniently written in terms of $Y_m$ it is natural to ask
for a dynamical equation for this variable. Re-expressing the
equation for ${\cal U}_m$ in terms of it we find \ba\label{Y}
-{\ddot Y}_m+2(\dot A+\frac{\ddot{\Phi}_0}{\dot{\Phi}_0}){\dot
Y}_m- 2{\ddot A}Y=m^2e^{-2A}Y_m. \ea Equations~(\ref{N})
and~(\ref{Y}) define completely the mass eigenvalue problem in the
longitudinal gauge~\cite{peloso}.

By an inspection of~(\ref{dyn}), and making use of the asymptotic
form of the background~(\ref{EOM}), one can conclude that the mass
squared are positive and continuous as $m>0$~\cite{Gio}. Because
of~(\ref{0}) we see that the eigenvector of zero mass can be
obtained by solving a first order equation ${\cal A}^-{\cal G}=0$
(the additional solutions of the system~(\ref{dyn}) do not satisfy
the constraint~(\ref{const})). The independent solutions can be
derived explicitly and are ${\cal G}=0,\cal Z$. In terms of the
original fields they read \ba\label{00}
\phi_0&=&\beta_1\frac{\dot\Phi_0}{a^2}-\beta_2\frac{\dot\Phi_0}{a^2}\int^y
a^2\\\no F_0&=&\beta_1\frac{2\dot
A}{a^2}+\beta_2\left(1-\frac{2\dot A}{a^2}\int^y a^2\right). \ea
The solution $\beta_2=0$ resembles the zero mode of the global
theory (the global theory admits a second zero mode solution,
$\phi_0=\dot\Phi_0\int^y dy\,1/\dot\Phi_0^2$, which is not
reproduced by the gravitational model. This mode is
nonnormalizable in the nongravitational model and thus it plays
no role in the dynamics), and it is instructive to observe that
$N_0(\beta_2=0)=\Delta\left(We^{-2A}\right)$. The latter
expression has to be compared with the vacuum energy of the
gravitational system, i.e. the brane tension, \ba
\rho[\Phi_0,A]=2\Delta\left(We^{4A}\right). \ea By
self-consistency, the latter must vanish (the effective 4d theory
is defined in flat space) so that the solution $\beta_2=0$ has a
divergent normalization $N_0$. We stress that the divergence of
the integral is entirely due to the mixing with the field $F$,
i.e. the $F_m^2$ term in the normalization~(\ref{N}).

Both the boundary and the divergent nature of $N_0$ hold for any
zero mode solution of the scalar system. To see this explicitly we
use equation~(\ref{Y}) to recast eq.~(\ref{N}) in the form
\ba\label{Nm} N_m=\frac{3M^3}{4}\int dy \left[-\frac{m^2}{\ddot
A}\left(\frac{Y_m}{a^2}\right)^2-\partial_y\left(\frac{Y_m\dot
Y_m}{a^2\ddot A}\right)\right].\ea By plugging the
solutions~(\ref{00}) into~(\ref{Nm}) we see that $N_0\propto
e^{-2A(\infty)}$ for any $\beta_{1,2}$, and therefore diverges. This
is tantamount to say that the zero mode decouples from the
effective theory. We conclude that the spectrum reduces to a continuum starting at zero mass.

\section{Scalar bound states}

The gravitational theory describes a drastically different
spectrum compared to the one of the globally symmetric model. The
zero mode predicted by the nongravitational theory disappears and
no brane perturbation is expected to mediate a long range force.
Nevertheless, the brane can still be excited at arbitrary small
scales since the continuum starts at zero momentum. In this
section we will see that this continuum forms a bound state that
appears as a resonant mode to a 4d observer residing on the brane.

Resonant states are found as the eigenvalue problem with complex momentum. They were first identified in the context of alpha decay
by Gamow by imposing asymptotic outgoing wave behavior on the
wavefunctions. An explicit expression of the resonant condition
for our system will be presented later on. For the moment let us
stress that the only
field manifesting a resonant behavior is $\phi$. 

To derive
an explicit expression for the dynamical equation for $\phi_m$ we
can differentiate~(\ref{Y}) and use~(\ref{const}). As anticipated
in section II this equation can be conveniently written as a
function of $f$, with $\phi=\dot\Phi_0 f$. The expression is
rather involved and reads:\ba\label{p}
\partial_y\left(\frac{\dot fe^{2A}}{1+\frac{m^2}{2\ddot A
}e^{-2A}}\right)=-2\ddot A fe^{2A}.\ea The equation thus obtained is gauge invariant. Indeed, eq.~(\ref{gi}) tells us that in the longitudinal gauge the field $\phi$ coincides, up to a factor, to the frame-independent fluctuation $2A'{\cal G}+\Phi_0^2{\cal U}$. The implications of eq.~(\ref{p}) are therefore physical and do not depend on the gauge choice.

An exact solution
of~(\ref{p}) cannot be found even for the simplest backgrounds,
but we can find an approximate solution by matching two
expressions for $f_m$, namely $f_<$, defined for $0<|y|<y_\ast$,
and $f_>$, defined for $|y|>y_\ast$. The distance $y_\ast$ is of
order the brane width $1/w$. For localized defects and $m\neq0$
the terms $m^2/\ddot A\gg1$ as soon as we move away from the
brane. Near the brane, on the other hand, the smoothness of the
background requires $\dot A\approx\dddot A\propto
\ddot\Phi\approx0$ and leads to a condition for $f_<$. Our approximate expressions are:
\ba\label{><} -\ddot f_>-\dot f_>(4\dot
A+2\frac{\ddot\Phi}{\dot\Phi})&=&m^2e^{-2A}f_>\\\no -\ddot
f_<-2\ddot Af_<&=&m^2e^{-2A}f_<.\ea

In the limit
$y_\ast\rightarrow0$, in which the brane appears infinitely thin
$A\approx-k|y|$, and assuming exponentially localized scalar defects
$\dot \Phi\sim e^{-2w|y|}$, eqs.~(\ref{><}) can be compactly
written as \ba\label{sim} -\ddot f+4(k+w)sign(y)\dot
f+4k\delta(y)f=m^2e^{2k|y|}f.\ea Because of the localization of
the defect the mixing between $\phi$ and the scalars $F,S$ reduces
to a brane effect and translates into the boundary condition at
$y=0$. The latter corresponds to a localized positive mass term
and will play a role in the determination of the resonant
condition. Notice that the boundary condition could have been
deduced from the would-be massless profile
$\phi_0=\dot\Phi_0e^{-2A}$ found in the previous section. As the
above derivation makes it clear, however, the simplified
eigenvalue problem is a reasonable approximation for massive modes
only.

The model described by the equation~(\ref{sim}) is a particular
example of a more general class of theories that will be
introduced in the following section. The system~(\ref{sim}) can be
used as a toy model for the description of our nontrivial set up,
having the great advantage of being exactly solvable. The
solution, up to a normalization, is \ba\label{appsol} f_m&=&e^{\nu
k|y|}\left(J_\nu\left(\frac{m}{k}e^{k|y|}\right)+\beta_m
Y_\nu\left(\frac{m}{k}e^{k|y|}\right)\right)\\\no
\nu&=&2\left(1+\frac{w}{k}\right),\ea where $J_\nu$ and $Y_\nu$
are the Bessel function of order $\nu$ of first and second kind
respectively. Imposing the boundary condition at the brane and
estimating $N_m$ using the normalization~(\ref{N}) (for $m\neq0$
the normalization is dominated by the $\phi$ integral) we find \ba
\beta_m&=&-\frac
{\frac{m}{k}J_{\nu-1}\left(\frac{m}{k}\right)-2J_\nu\left(\frac{m}{k}\right)}
{\frac{m}{k}Y_{\nu-1}\left(\frac{m}{k}\right)-2Y_\nu\left(\frac{m}{k}\right)}\\\no
f_m&\propto&\sqrt{\frac{m}{k}}\frac{e^{\nu
k|y|}}{\sqrt{1+\beta_m^2}}\left(J_\nu\left(\frac{m}{k}e^{k|y|}\right)+\beta_m
Y_\nu\left(\frac{m}{k}e^{k|y|}\right)\right)\ea where the
proportionality factors are numbers depending on the scale
entering $\Phi_0$ (and the brane tension $\Delta W$), which we can
simply ignore.

A plot of the spectral density at $y=0$, i.e. $\phi_m^2(0)$,
reveals the presence of a continuum starting at zero; this is very
suppressed up to an energy of the order $m^2\sim kw$, where it
develops a very sharp peak near the pole of $\beta_m$, and finally
stabilizes at a scale $m>w$. The peak is the remnant of the delta
function localized zero mode density of the global theory which,
remarkably, has been shifted to a nonzero mass by gravity. A
similar result has been previously found in~\cite{STZ}. In order to understand more deeply the
nature of this peak we isolate it by demanding for the appearance
of resonant modes.

Imposing outgoing wave boundary conditions on~(\ref{appsol})
(setting $\beta_m=i$) we find the explicit form of the resonant
state of our model~(\ref{sim}) \ba\label{res}
\phi_{R}=\dot\Phi_0e^{\nu
k|y|}H^{(1)}_\nu\left(\frac{m}{k}e^{k|y|}\right),\ea where
$H^{(1)}_{\nu}=J_{\nu}+iY_{\nu}$ is the Hankel function of the
first kind. The boundary condition on the brane fixes completely
the eigenvalue at a complex $m=m_R-i\Gamma_R/2$. If the pole is
located in the lower half plane, i.e. $\Gamma_R>0$, an
asymptotically outgoing wave manifests the characteristic
exponential decay in time (and, typically, an exponential
divergence in space) and the quantities $m_R$ and $\Gamma_R$ can
be identified as the mass and the width of the resonance,
respectively. The resonant condition for our system finally reads:
\ba~\label{Res}
\frac{m}{k}\frac{H^{(1)}_{\nu-1}\left(\frac{m}{k}\right)}
{H^{(1)}_{\nu}\left(\frac{m}{k}\right)}=2.\ea This condition
admits a single solution at a scale $m_R^2\sim kw$, corresponding
to a complex pole in the scalar propagator. A similar procedure
applied to $F$ (or, equivalently, to ${\cal G}$ and ${\cal U}$) provides no solution.

An explicit analytical identification of the root
$m=m_R-i\Gamma_R/2$ is not possible since, strictly speaking, the
Bessel functions cannot be approximated by an expansion in either
the small or large argument limit for $m^2\sim kw$. Nevertheless a
numerical investigation shows that for sufficiently large $\nu$
the approximation of small argument works quite well. This
approximation allows us to obtain expressions similar to those
of~\cite{DRT}\footnote{The approximation $x\ll\sqrt{\nu}$ was
natural in the framework studied by~\cite{DRT} since they
considered a case analog to $w\ll k$, which in our framework is
not a reliable limit.}:\ba~\label{approx}
x\frac{H^{(1)}_{\nu-1}\left(x\right)}
{H^{(1)}_{\nu}\left(x\right)}\approx
x\frac{Y_{\nu-1}}{Y_{\nu}}\left[1-ix\frac{J_{\nu-1}}{Y_{\nu-1}}\right],\ea
where for brevity $x=m/w$, that will be used to estimate the pole
in the following. The reliability of the small argument
approximation should not come as a surprise, since comparing the
small argument limit of~(\ref{Res}) with the explicit expression
for $\beta_m$ we see that the pole in the Green's function
approaches the bump found in the spectral density $\phi_m^2(0)$.
The same approximation implies that for $k|y|<1$ the resonant
profile~(\ref{res}) acquires approximately the same $y$ dependence
of the zero mode $\sim\dot\Phi_0$, while outside the brane the
resonance becomes exponentially divergent because of the positive
width.

We can now find an approximate expression for the resonant
propagator. Because of the mixing among $\phi,F$, and $S$, the
scalars' Green's function is a three by three matrix where each of
the entries can be formally written in the canonical form
$$G_{ab}(p^2,y,y')\propto \int
dm\,\frac{g_m^{(a)}(y)g_m^{(b)}(y')}{p^2-m^2},$$ with
$g^{(a,b)}_m=\phi_m,F_m,S_m$ (see~\cite{STZ} for an explicit
expression of this matrix in the light-cone gauge). Under the simplifying assumption of an
exponentially localized defect, the mixing between the scalar
states becomes a pure brane effect and the Green's function
$G_{\phi\phi}$ is completely determined by the eigenmode $\phi_p$
and its boundary conditions. Following an analog procedure as that
used in section II we obtain an expression for the brane to brane
propagator as: \ba
G_{\phi\phi}(p^2,0,0)=\frac{1}{\frac{\dot\phi_p(0)}{\phi_p(0)}-2k}.\ea
Setting $\phi_p=\phi_R$ in the above expression and
using~(\ref{approx}) we are able to isolate a pole
\ba~\label{propR}
G_{\phi\phi}(p^2,0,0)&\sim&\frac{1}{p\frac{H^{(1)}_{\nu-1}\left(\frac{p}{k}\right)}
{H^{(1)}_{\nu}\left(\frac{p}{k}\right)}-2k}\\\no
&\approx&\frac{2k(\nu-1)}{p^2-m_R^2+ip\,\Gamma_R(p)} \ea where we
have defined: \ba~\label{resmass}
\left(\frac{m_R}{k}\right)^2&\approx& 4(\nu-1)\\\no
\frac{\Gamma_R(m_R)}{m_R}&\approx&
\frac{\pi/4}{\Gamma(\nu-1)\Gamma(\nu)}\left(\frac{m_R}{2k}\right)^{2(\nu-1)}.\ea
As already emphasized, eqs.~(\ref{resmass}) are only approximate,
though reliable, expressions of the pole. Eqs.~(\ref{propR})
and~(\ref{resmass}) represent the main results of the paper.

The resonant propagator should be compared with~(\ref{propNGB}).
As gravity decouples we have $m_R-i\Gamma_R/2\rightarrow0$ as
$k/w\rightarrow0$, and the resonance exchange mimics the zero mode
of the global theory ensuring continuity of the observable as
gravity decouples. The ratio $\Gamma_R/m_R\propto(e^2/\nu)^\nu$ is
exponentially small and can be neglected in the limit $k\ll w$.
The resonance is thus a stable state for all practical purposes
and mediates a potential between static brane sources of the form
\ba V_R\propto\frac{1}{r}e^{-m_Rr},\ea where, in the limit
$w/k\gg1$, $m_R^2\approx8wk\ll w^2$. The residue in~(\ref{propR})
is the $4w$ factor found in~(\ref{propNGB}) except for small
O$(k/w)$ corrections. This means that the scale determining the
effective coupling of the resonance to brane localized currents is
approximately the same as the zero mode of the global theory.

For $k\ll w$ the resonance represents the only relevant brane
fluctuation below the cut-off $w$, as the spectral function
$\phi_m^2(0)$ suggests. We can appreciate the negligible impact on
the effective theory of the continuum $m\ll m_R$ by estimating the
Yukawa potential they mediate on the brane. At 4d distances $r$
large compared to the curvature $kr\gg1$ these are expected to
play a role. Expanding for small arguments our approximate
solutions we have
$f_m(0)\propto\sqrt{x}xJ_{\nu-1}(x)\propto\sqrt{x}x^\nu$, with
$x=m/k$, and the potential behaves as\ba\label{pot}
V_{cont}\sim\int dm \frac{\phi_m^2(0)}{r}e^{-mr}\propto
\frac{1}{\nu r}\left(\frac{1}{kr}\right)^{2\nu+2}\ea for $kr\gg1$.
In the limit $\nu\gg1$ the effect of the continuum below the
resonant peak is highly suppressed. An analogous suppression holds
for the continuum in the range $m_R<m<w$. Notice
that the continuum becomes relevant at $m\sim w$, rather than at
$2w$ as in the global theory.

Having established the dominant role of the resonance in the low
energy brane to brane $\Phi$ exchanges we conclude that, provided
$k\ll w$, the low energy dynamics acquires a dependence on the new scale $m_R^2\sim kw$. For $w^2\gg p^2\gg m_R^2$ the
resonance represents the only dynamical degree of freedom of the
brane. In this regime the global description is an approximation
of the full theory up to O$(k/w)$ corrections, and gravity can be seen as a {\it{weak gauging}} of the Poincare group. More interesting is
the opposite limit. For $p^2\ll m_R^2$ no brane excitation can be
significantly produced and the effect of the brane fluctuation on
the physics localized on the brane is totally encoded in
nonrenormalizable operators suppressed by the scale $m_R$.

Let us briefly comment on the generalization to other gauge
choices. In the axial gauge ($\delta g_{A5}=0$) the
constraint~(\ref{const}) simplifies to $\dot
F=-2\dot\Phi_0\phi/3$, but the dynamical equation for $Y_m$ can no
longer be used to derive a resonant condition for $\phi_m$. It is more convenient to derive a third order
differential equation for $f=\phi/\dot\Phi_0$ from the dynamical
condition for $\cal G$ with the help of~(\ref{const}). This
equation was found in~\cite{DeWolfe}, but here it will not be
reproduced for brevity. The latter leads to a resonant condition
of the same form as~(\ref{Res}) (with $\nu\rightarrow\nu-1$ and
$2\rightarrow4$), and consistently formulas similar
to~(\ref{propR}),~(\ref{resmass}), and~(\ref{pot}) can be derived.
On the other hand, the same line of reasoning applied to the
solution found in~\cite{STZ} using the light-cone gauge reveals
the presence of a resonant mode described by an equation similar
to~(\ref{Res}), but with $\nu\rightarrow1$. In this case the solution has
mass and width of comparable magnitude, and both proportional to
the curvature scale. In addition, in the light-cone gauge the authors~\cite{STZ} found that the
potential mediated by the continuum of light modes in the regime
$kr\gg1$ is suppressed by $1/(kr)^2$  with respect to the long
range $1/r$, while in both the
longitudinal and the axial gauge the suppression is much more
significant (at least $1/(kr)^{2\nu}$).

One should not be worried in finding different predictions for the
resonant pole: the resonance appears in the gauge-dependent propagator of the field $\phi$. Only in the longitudinal gauge the dynamical equation for $\phi$ provides gauge-invariant information.

\section{An exactly solvable model}

The appearance of resonant modes in place of the discrete states
of the global theory resembles the situation considered in
~\cite{DRT}. In that paper the authors studied the effect of the
noncompact Randall-Sundrum warping on chiral fermions localized
with the mechanism described in~\cite{RubShap}. They found that if
we supply the fermionic field with a bulk mass, then the massless
localized state of the nongravitational theory disappears and the
continuum generates a resonant mode with mass proportional to the
small bulk mass and an exponentially suppressed width. We will now
show that a similar situation is realized in the spin-0 sector of
the gravitational theory considered in the present paper.

Consider a bulk scalar field $\psi$ fluctuating in a gravitational
background \ba {\cal L} = \sqrt{-g}\left(
g^{AB}\partial_A\psi\partial_B\psi-m_5^2\psi^2)\right). \ea In
order to localize light modes on the brane we introduce an
attractive delta function mass term. Let us first discuss the theory in flat space and define
$m_5^2=4(w^2-w_b\delta(y))$. The mass eigenvalue problem for the
scalar wavefunctions $\psi_m$ can be written as: \ba\label{sm}
\int dy\,\psi_m(y)\psi_{m'}(y)&=&\delta_{m,m'}\\\no
-\ddot\psi_m+4(w^2-w_b\delta(y))\psi_m&=&m^2\psi_m, \ea where
$w_b$ is assumed to be a free parameter of order $w$. The
system~(\ref{sm}) admits a continuum starting at the threshold
$2w$ and a single localized mode. The normalizable wavefunction
for the latter is $\psi=e^{-\mu |y|}$ and has a mass squared
$4w^2-\mu^2$. The parameter $\mu$ is determined by the boundary
conditions as $\dot\psi(0)=-2w_b\psi(0)$, which gives $\mu=2w_b$.
We see that normalizability enforces $w_b\geq0$ while absence of
tachions $w_b\leq w$ (the delta function potential cannot be too
attractive).

As gravity is switched on eq.~(\ref{sm}) receive corrections both
in the potential and the kinetic term. For a geometry of the
Randall-Sundrum form $ds^2=e^{-2k|y|}\eta_{\mu\nu}dx^\mu
dx^\nu-dy^2$ we have
\ba\label{gsm} \int
dy\,e^{-2k|y|}\psi_m(y)\psi_{m'}(y)&=&\delta_{m,m'}\\\no
-\ddot\psi_m+4k\,sign(y)\dot\psi_m+4(\bar w^2-\bar
w_b\delta(y))\psi_m&=&m^2e^{2k|y|}\psi_m \ea where now both $\bar
w$ and $\bar w_b$ may contain curvature corrections of order $k/w$
with respect to $w$ and $w_b$ respectively. Under the assumption
$\bar w^2> \bar w_b(\bar w_b+2\bar w)$ the spectrum has no
tachionic modes, while the requirement $\bar w^2= \bar w_b(\bar
w_b+2\bar w)$ is the necessary condition for the existence of a
localized massless mode.

The model~(\ref{gsm}) typically admits resonant modes. Imposing outgoing boundary conditions we find:
\ba\label{Toy}
\frac{m}{k}\frac{H^{(1)}_{\nu-1}\left(\frac{m}{k}\right)}
{H^{(1)}_{\nu}\left(\frac{m}{k}\right)}&=&\left(\nu-2-2\frac{\bar
w_b}{k}\right)\\\no \nu&=&2\sqrt{1+\frac{\bar w^2}{k^2}}. \ea Thin
resonances exist only if the right hand side of the first
equation~(\ref{Toy}) is positive and, not surprisingly, if $\bar
w_b\neq0$. The former requirement is the same condition ensuring
absence of tachionic excitations. Proceeding as in the previous
section we approximately find a solution for $m=m_R-i\Gamma_R/2$,
where \ba\label{toy} \left(\frac{m_R}{k}\right)^2&\approx&
2(\nu-1)\left(\nu-2-2\frac{\bar w_b}{k}\right)\\\no
\frac{\Gamma_R(m_R)}{m_R}&\approx&
\frac{\pi/4}{\Gamma(\nu-1)\Gamma(\nu)}\left(\frac{m_R}{2k}\right)^{2(\nu-1)}.\ea

The theory analyzed in the text is found to be a particular
example of~(\ref{gsm}) with $\bar w^2=w(w+2k)$ and $\bar w_b=w-k$
(see eq.~(\ref{sim}) in the text). Despite the involved expression
for the $\phi_m$ norm obtained in section III, it is easy to
verify that the $F^2$ integral in eq.~(\ref{N}) is convergent for
any $m\neq0$ and, thus, the normalization condition for the
fluctuations $\phi_m$ is effectively of the same form
as~(\ref{gsm}). As gravity decouples ($k=0$) $\bar w,\bar
w_b\rightarrow w$ and the model predicts a localized massless
mode\footnote{The kink background~(\ref{kink}) reproduces the
above potential in the thin brane limit $w\rightarrow\infty$. For
an observer at a distance $y>1/w$, where $1/w$ characterizes the
thickness of the defect, the background can be simplified by the
approximation $\tanh(wy)\simeq sign(y)$. From the identities
$sign(y)^2=1$ and $\partial_y sign(y)=2\delta(y)$ we derive
$\dddot\Phi_0/\dot\Phi_0\simeq-2w^2(2\delta(y)/w-2)$.}. Because of
the mixing with gravity, an O$(k/w)$ correction to the mass term
is induced and the mode disappears from the spectrum leaving a
resonance in its place.

Using this simplified picture one can also deduce the fate of a
possibly massive localized mode. An inspection of~(\ref{toy})
shows that, in the limit $k\ll w$ and for $\bar w_b=$ O$(\bar w)$,
a resonance still appears at $m_R\sim w$ but quickly becomes wider
as its mass increases. For example the choice $\bar w^2=w(w+2k)$
and $\bar w_b=w_b$ leads to the prediction of a broad resonance
with mass $m_R^2\approx 8w^2(1-w_b/w)$, that should be compared
with the flat space-time result $4w^2(1-w_b^2/w^2)$. Although this
choice seems quite arbitrary, one can verify with techniques similar
to those used in the bulk of the paper that it provides a good
approximation of the next to higher state of the kink background.

\section{Summary and Conclusion}

We analyzed in detail the scalar fluctuations around nontrivial
gravitational backgrounds triggered by the space-dependent vacuum
expectation value of a scalar field $\Phi$. Assuming the scalar
defect localizes light states around a particular region of the
noncompact extra dimension, we focused on the role of gravity on
the brane-localized physics.

In the globally symmetric version of the model, the dimensionally
reduced theory of the $\Phi$ fluctuations describes a localized
zero mode, possibly discrete eigenvalues, and a continuum
interacting with the physics residing on the brane. Including the
gravitational field no scalar bound state persists, but the continuum of
physical spin-0 fields is found to generate massive resonances in
the 2-point function of the scalar $\Phi$. Because of the frame-dependent nature of the latter, the resonances turn
out to be gauge dependent entities. We estracted frame-independent predictions by working in the longitudinal gauge. 

Under some reasonable
assumptions we realized that the main features of the coupled
scalar-gravity system can be described by an exactly calculable
model. We found an
explicit expression for the resonant propagator and showed that
the heavier the resonance the wider its peak. At distances much
larger than the brane width the only possibly relevant brane
deformation is the approximately stable lightest resonance. Its couplings to brane localized currents are those of the
zero mode of the global theory except for small corrections, but its nonvanishing mass introduces new scales in the IR theory. We have shown that this new scale depends on 
the brane width.

\acknowledgments
This work is partially supported by MIUR
and the RTN European Program MRTN-CT-2004-503369.


 \end{document}